\documentclass[a4 paper, 11pt]{article}
\usepackage{jheppub} % for details on the use of the package, please
\usepackage{subcaption}
\usepackage{xcolor}
\usepackage[T1]{fontenc} % if needed
\usepackage{xspace}
\usepackage{multirow}
\usepackage{hyperref}
\usepackage{slashed}
\usepackage{amsmath}
\usepackage{tikz-feynman}
\usepackage{epstopdf}

\usepackage[utf8]{inputenc}

\tikzfeynmanset{compat=1.1.0}

\def\GeV{\ifmmode {\mathrm{\ Ge\kern -0.1em V}}\else \textrm{Ge\kern -0.1em V}\fi}%
\def\GeV{\ifmmode {\mathrm{\ Ge\kern -0.1em V}}\else \textrm{Ge\kern -0.1em V}\fi}%

\newcommand{\startappendix}{
\setcounter{section}{0}
\renewcommand{\thesection}{\Alph{section}}
\renewcommand{\theequation}{\Alph{section}.\arabic{equation}}}

\newcommand{\Appendix}[1]{
\refstepcounter{section}
\begin{flushleft}
{\Large\bf Appendix: #1}
\end{flushleft}}

\title{\boldmath Gravitational waves and dark matter from classical scale invariance}

\author{Valentin V. Khoze and}
\author{Daniel L. Milne}

\affiliation{IPPP, Department of Physics, Durham University, Durham DH1 3LE, UK}

\emailAdd{valya.khoze@durham.ac.uk}
\emailAdd{daniel.l.milne@durham.ac.uk}

\abstract{
In this paper we consider a minimal classically conformal U(1) model of fermionic dark matter. We calculate the one loop effective potential which generates the mass scale quantum mechanically via dimensional transmutaion in the spirit of Gildener and Weinberg, and examine the effects of the new dark sector on the Standard Model Higgs as well as how the dark fermions receive a mass and can produce the observed relic abundance. We then consider constraints on the model coming from collider and direct detection experiments before calculating the thermal effects on the potential in the early universe. We examine the nature and conditions for a strongly first order phase transition in our model and calculate the associated gravitational wave signal and compare to the sensitivities of current and proposed experiments.
}

\begin{document}
\preprint{}

\maketitle
\flushbottom
%\newpage

%%%%%%%%%%%%%%%%%%%%%%%%%%%%%%%%%%%%%%%%%%%%%%%

\section{\label{Sec:Intro}Introduction}

After the discovery of the Higgs boson in 2012 by the ATLAS and CMS experiments \cite{ATLAS:2012yve,CMS:2012qbp}, the Standard Model was complete and potentially valid all the way up to the Planck scale. However it is known that there are problems with the Standard Model and it cannot be a complete description of reality. One of these problems is the so-called hierarchy problem, the fact that the Higgs mass is unstable against quantum corrections; there is also the issue that we require dark matter in order to explain the observed density of the universe and new physics is also needed to explain the baryon asymmetry of the universe. The model proposed in this paper goes some way to addressing each of these concerns and would also have experimentally observable consequences.

In the Standard Model, the Higgs vev is introduced at tree level, but in the 1970s Coleman and Weinberg showed that it was possible for the tree level potential to have its minimum at the origin and still develop a minimum away from the origin at loop level \cite{Coleman:1973jx}. This idea was known as dimensional transmutation as we trade the dimensionful parameter of the higgs quadratic term for the dimensionless parameter of the scalar coupling. As this also removes the dimension-2 operator in the standard model, it removes the source of corrections which depend quadratically on the UV cutoff and hence is believed by many, including the present authors, to solve the Hierarchy problem\footnote{\label{footnote:HPDebate}Though this point is still open for debate and is far from being settled, see e.g. \cite{MarquesTavares:2013szc}.}\cite{Bardeen:1995kv,Hempfling:1996ht}. However due to the relationship induced between couplings in the dimensional transmutation it has been known since the 90s that the Standard Model Higgs is too heavy to come from a Coleman-Weinberg theory and one then applies the Coleman-Weinberg mechanism in a hidden sector~\cite{Hempfling:1996ht,Meissner:2006zh,Chang:2007ki,Foot:2007iy,Englert:2013gz}
 coupled to the Standard Model for example via a Higgs portal interaction.
 
It has also been known for many years that some additional matter to that observed in the universe is needed to explain many observations. The first piece of evidence was seen in galactic rotation curves \cite{Rubin:1970zza} and there have since been many other pieces of evidence, such as observations of galactic collisions \cite{Markevitch:2003at} and data from the CMB experiment \cite{Planck:2015fie}, which support the hypothesis of particle dark matter \cite{Bertone:2004pz}. A popular model of dark matter is that the dark matter particle is charged under some new gauge group while being a singlet under the standard model gauge group and conversely all Standard Model particles are singlets under the dark gauge group. The number of particles in the dark sector (those charged under the new gauge group) varies heavily from model to model, some are very minimal including only one particle, while some contain multiple vectors, scalars and fermions. The dark sectors then 'communicate' with the Standard Model through either a Higgs portal or kinetic gauge mixing. See \cite{
Cheung:2007ut,
Hambye:2008bq,
Lindner:2011it,
Baek:2012se,
Hambye:2013dgv,
Carone:2013wla,
Khoze:2014xha,
Khoze:2014woa,
Harris:2014hga,
%Harris:2015kda,
%Abercrombie:2015wmb,
Alves:2015mua,
Rodejohann:2015lca,
Karam:2015jta,
Karam:2016rsz,
Khoze:2016zfi,
Bauer:2018egk,
Foldenauer:2018zrz,
Nomura:2020zlm,
Baouche:2021wwa,
Tapadar:2021kgw,
Dasgupta:2021dnl,
Barman:2022njh,
Ghorbani:2015xvz} for examples.

Finally, it was demonstrated by Sakharov that in order to generate the required baryon asymmetry of the universe, thermal equilibrium must be violated in the early universe (amongst other requirements)~\cite{Sakharov:1967dj}. This requirement can be satisfied if there is a first order phase transition in the early universe. However it is known that in the Standard Model the electroweak phase transition is second-order/crossover, but this can be modified with the addition of new physics to create a first-order phase transition and go some way to generating the baryon asymmetry of the universe e.g. \cite{Croon:2018kqn,Li:2020eun,Athron:2019teq}. It is also well known that a first order phase transition can lead to the production of gravitational waves. There were several papers in the late 1980s - early 1990s which first calculated this possibility \cite{Hogan:1986qda,Kosowsky:1992rz,Kosowsky:1991ua,Kamionkowski:1993fg} and it is now a common consideration in dark matter models \cite{Schwaller:2015tja,Jaeckel:2016jlh,Mohamadnejad:2019vzg,Addazi:2017gpt,Breitbach:2018ddu,
Croon:2018erz,Croon:2019iuh,Ellis:2018mja,Ellis:2019oqb,Chala:2019rfk,Ghosh:2020ipy,Azatov:2021ifm,Brdar:2018num,Hall:2019ank,Huang:2020crf}.

In this paper, we attempt to simultaneously alleviate these various shortcomings by augmenting the Standard Model with a dark sector and imposing classical scale invariance. In this way we can generate the required dark matter abundance whilst simultaneously solving the hierarchy problem through the demand of conformal symmetry at the the classical level (see footnote \ref{footnote:HPDebate}). With this new model it will also be possible to generate a strongly first order electroweak phase transition, thus aiding in the generation of the baryon asymmetry (although it should be noted that we will not seek to satisfy all of Sakharov's conditions). 

The structure of the paper is as follows; in Section~\ref{sec:model} we introduce the model and derive the symmetry breaking. In Section~\ref{sec:expCon} we calculate the relic abundance of the model and examine in which areas of parameter space we can reproduce the observed relic abundance as well as imposing constraints arising from collider searches for new particles and direct detection experiments looking for dark matter. In Section~\ref{sec:theCon} we examine the theoretical constraints on the model such as ensuring that the vacuum is stable, we respect perturbative unitarity and that the model is perturbative. Finally, in Section~\ref{sec:gravWaves} we calculate the potential in the early universe before looking at which areas of phase space give rise to a strongly first order phase transition and the associated gravitational wave signal before we conclude in Section~\ref{sec:conc}.
\medskip

\section{\label{sec:model}Model and the Radiative Generation of Scale}

We introduce a classically scale invariant model with a dark sector charged under a new $U(1)$ symmetry and coupled to the standard model through a Higgs portal coupling. Our model is similar to that considered in \cite{Kim:2019ogz}, although we extend their model by allowing our fermions to have different masses (as well as in later sections looking at the model in the early universe by investigating the phase transition and a production of gravitational waves). The model is given by
\begin{equation}
\mathcal{L}=\mathcal{L}_{\rm{SM}}+\mathcal{L}_{kin}+\mathcal{L}_{Y}-V_0\left(H,S\right)
\end{equation}
where $\mathcal{L}_{SM}$ is the Standard Model lagrangian without the Higgs potential, $\mathcal{L}_{kin}$ is the kinetic terms for the new fields\footnote{Note that the term $\epsilon F_{\mu\nu}F'^{\mu\nu}$, where $F_{\mu\nu}$ is the U(1)$_{Y}$ field strength tensor, is also allowed by gauge invariance but we neglect this term in light of strong collider constraints\cite{Hook:2010tw,Curtin:2014cca} and leave it for future work.} :
\begin{equation}
\mathcal{L}_{kin}=\lvert D_{\mu} S\rvert^{2}-\frac{1}{4}F'_{\mu\nu}F'^{\mu\nu}+\overline{\chi}^{a}_{L}\slashed{D}\chi^{a}_{L}+\overline{\chi}^{a}_{R}\slashed{D}_{\mu}\chi^{a}_{R}.
\end{equation}
All new particles, $S, \chi, A'_{\mu}$ are singlets under the standard model gauge group, $G_{\rm{SM}}=SU(3)_{C}\times SU(2)_{L}\times U(1)_{Y}$ and all SM particles are singlets under the new gauge group. The charges of the new particles are given in Table~\ref{table:charges}. $F'_{\mu\nu}$ is the field strength tensor associated with the gauge boson of the new $U\left(1\right)$ and there is implied summation over repeated indices.  

\begin{table}
\centering
\begin{tabular}{|c|c|c|c|c|c|}
\hline
Field & $S$ & $\chi^{1}_{L}$ & $\chi^{1}_{R}$ & $\chi^{2}_{L}$ & $\chi^{2}_{R}$ \\
\hline
$U(1)_{D}$ & 1 &  $\frac{1}{2}$ & $-\frac{1}{2}$ & $-\frac{1}{2}$ & $\frac{1}{2}$ \\
\hline
\end{tabular}
\caption{The charges of the dark sector particles under the new $U(1)_{D}$ symmetry. Note that this assignment of charges renders the theory anomaly-free.}
\label{table:charges}
\end{table}

$\mathcal{L}_{Y}$ is the Yukawa coupling of the dark sector:
\begin{equation}
\mathcal{L}_{Y}=y_{1,D}\overline{\chi}^{1}_{L}S\chi^{1}_{R}+y_{2,D}\overline{\chi}^{2}_{L}S\chi^{2}_{R}+h.c.
\end{equation}
The tree-level potential for scalar fields of the new theory is given by:
\begin{equation}
\label{eq:potential}
V_0\left(H,\,S\right)=\lambda_{H}\left(H^{\dagger}H\right)^{2}+\lambda_{S}\left(S^{*}S\right)^{2}-\lambda_{P}\left(H^{\dagger}H\right)\left(S^{*}S\right).
\end{equation}
Note that we require $\lambda_{P}>0$ to create a true minimum away from the origin. Working in the unitary gauge where we can write
\begin{align}
   H &=\frac{1}{\sqrt{2}} \begin{bmatrix}
           0\\
           h
         \end{bmatrix}
\quad
   S =\frac{s}{\sqrt{2}} 
\end{align}
where $h$ and $s$ are real; then the classical scalar potential may be written as
\begin{equation}
V_0\left(h,\,s\right)=\frac{\lambda_{H}}{4}h^{4}+\frac{\lambda_{S}}{4}s^{4}-\frac{\lambda_{P}}{4}h^{2}s^{2}.
\label{eq:V0hs0}
\end{equation}

This potential is classically scale invariant in the sense that it does not contain any dimensionful parameters at tree level, and all scales have to be generated radiatively by taking into account quantum effects. 
Symmetry breaking in a classically scale invariant model was first considered in \cite{Coleman:1973jx}. In models such as ours many authors consider $\lambda_{P}$ to be small so that the backreaction of the Standard Model on the dark sector is negligible and one can treat the dark sector in the original Coleman-Weinberg formalism (see e.g. \cite{Hempfling:1996ht, Englert:2013gz,Oda:2015gna}) however a more general formalism was later developed by Gildener and S. Weinberg in \cite{Gildener:1976ih} to deal with theories of multiple scalars and it is this formalism we shall follow here in order not be restricted in our choice of parameters.

\subsection{The Coleman-Weinberg Approach}

The seminal Coleman-Weinberg paper \cite{Coleman:1973jx} considered a simple model of a {\it single} classically massless scalar field $\phi$ interacting with gauge fields. The effective potential for $\phi$ at 1-loop level is given by (for the case where $\phi$ is coupled to $SU(2)$ gauge fields with the gauge coupling $g$, see e.g. \cite{Khoze:2016zfi}),
\begin{equation}
V_{CW}(\phi)=\frac{\lambda_\phi(\mu)}{4}\phi^4 +\frac{9 }{1024\,\pi^2}\, g^4(\mu)
\,\phi^4\left(\log\frac{\phi^2}{\mu^2}-\frac{25}{6}\right).
\label{eq:VCWsimple}
\end{equation}
Here $\mu$ is the RG scale and we are keeping 1-loop corrections arising from interactions of
$\phi$ with the SU(2) gauge bosons in the hidden sector, but neglecting the loops of $\phi$.
To justify this, the scalar self-coupling 
$\lambda_\phi \phi^4/4$ 
at the relevant scale $\mu =\langle\phi\rangle$ is assumed to be small, $\lambda_\phi \sim g^4 \ll 1$ relative to the 
1-loop gauge fields contributions to $V_{CW}(\phi)$.
This can be achieved as follows: in a theory where $\lambda_\phi$ has a positive slope, we start at a relatively high scale where $\lambda_\phi$ is positive and move toward the infrared until approach the value of $\mu$ where $\lambda_\phi(\mu)$ becomes small and is about to cross zero. This is the Coleman-Weinberg scale where the potential \eqref{eq:VCWsimple} develops a non-trivial minimum and $\phi$ generates a non-vanishing vev.
Minimising $V_{CW}(\phi)$ at $\mu=\left<\phi\right>$ gives:
\begin{equation}
\lambda_\phi \,= \, \frac{33}{256\,\pi^2} \,g^4 
\qquad {\rm at} \quad \mu=\langle \phi\rangle,
\label{eq:cwmsbar-PSU2}
\end{equation}
which is consistent with neglecting the scalar loop contributions to \eqref{eq:VCWsimple} in the first place.

This approach can be generalised to scale-invariant models with multiple scalar fields if one assumes that the portal couplings of additional scalars to the CW scalar field $\phi$ are small. In this case one uses a two-step process, where one first generates the vev for the CW scalar $\langle \phi \rangle$ and then transmits this vev via portal couplings to the remaining scalars as explained in \cite{Hempfling:1996ht, Englert:2013gz}.

In this paper we will follow instead a more general formalism of Gildener and Weinberg in \cite{Gildener:1976ih} where no assumptions need to be made about the size of the portal couplings to additional scalars to minimise their backreaction to the CW scalar vev.

\subsection{The Gildener-Weinberg Method}

In this approach we do not need to choose which of the available scalar fields plays the role of the CW scalar that develops the vev first. Here all available scalars are treated democratically.
To apply the method~\cite{Gildener:1976ih}  to  our two-scalar field model \eqref{eq:V0hs0} we write the scalars $h$ and $s$ in the form,
\begin{equation}
h=n_{1}\phi \quad s=n_{2}\phi,
\label{eq:shphidef}
\end{equation}
where $\textbf{n}=(n_1,n_2)$ is a certain 2D unit vector that will be determined below, and $\phi$ is the analogue of the CW scalar field whose vev is arbitrary at the classical level but fixed by the inclusion of quantum effects in the effective potential.

The classical potential \eqref{eq:V0hs0}  then reads
\begin{equation}
V_{0}\left(h,\,s\right)=\phi^{4}\left(\frac{\lambda_{H}}{4}n_{1}^{4}+\frac{\lambda_{S}}{4}n_{2}^{4}-\frac{\lambda_{P}}{4}n_{1}^{2}n_{2}^{2}\right).
\label{eq:phi4GW}
\end{equation}
The main idea of the 
Gildener-Weinberg approach is that, as in the the Coleman-Weinberg approach, there is a single classically flat direction in the scalar fields space, 
given by the ray $(h,s)= \textbf{n} \phi$, where $\textbf{n}$ is the unit vector specifying this ray, and $\langle \phi \rangle$ is arbitrary at tree-level. 
Specifically, this means that the tree-level $\phi^4$ vertex in \eqref{eq:phi4GW} is set to zero, or more precisely made parametrically smaller than the leading order 1-loop effects, in complete analogy with \eqref{eq:cwmsbar-PSU2},
\begin{equation}
\frac{\lambda_{H}}{4}n_{1}^{4}+\frac{\lambda_{S}}{4}n_{2}^{4}-\frac{\lambda_{P}}{4}n_{1}^{2}n_{2}^{2}  \, \ll\,  1.
\label{eq:phi4GW2} 
\end{equation}
The remaining directions in the scalar fields space are lifted already by the classical potential.
Thus we search for the minimum (or more precisely the classical flat direction) of $V_{0}$ by solving $\frac{\partial V_{0}}{\partial n_{i}}=0$ along with setting the right hand side of \eqref{eq:phi4GW2} to zero. This leads to the constraints:
\begin{gather}
\label{eq:GWConditions}
\lambda_{H}n_{1}^{2}-\frac{\lambda_{P}}{2}n_{2}^{2}=0\\
\lambda_{S}n_{2}^{2}-\frac{\lambda_{P}}{2}n_{1}^{2}=0\\
\label{eq:GWConditions3}
\frac{\lambda_{H}}{4}n_{1}^{4}+\frac{\lambda_{S}}{4}n_{2}^{4}-\frac{\lambda_{P}}{4}n_{1}^{2}n_{2}^{2}=0.
\end{gather}
In quantum theory, these equations are supposed to be satisfied at a certain value the RG scale $\mu=\Lambda_{GW}$, which is the dimensional transmutation scale responsible for the generation of the vev of the scalar field $\phi$.\footnote{In general, there is no guarantee that in every scalar theory one can find a point where these linear combinations of the couplings can be simultaneously set to zero, just like in the simple Coleman-Weinberg model there was no automatic guarantee that there was a scale at which the quartic coupling $\lambda_\phi$ was $\ll 1$. However for the families of the theories where the RG running allows these conditions to be satisfied, the value of the RG scale where it happens serves as the definition of the GW scale $\Lambda_{GW}$.}

The solution of the above equations determines the unit vector $\textbf{n}$ in terms of the couplings of the model,
\begin{gather}
\label{eq:flatDirection}
n_{1}^{2}=\frac{\lambda_{P}}{\lambda_{P}+2\lambda_{H}}\\
n_{2}^{2}=\frac{2\lambda_{H}}{\lambda_{P}+2\lambda_{H}}.
\end{gather}
We now expand the fields about their minima, writing $h=wn_{1}+\tilde{h},\,s=wn_{2}+\tilde{s}$ where $w=\langle \phi\rangle$ is a classically flat direction.\footnote{The value of $w$ will be stabilised below by the inclusion of quantum effects.} This leads to the mass matrix:
\begin{equation}
\label{eq:massMatrix}
M^{2}=w^{2}\begin{bmatrix}
2\lambda_{H}n_{1}^{2} &-\lambda_{P}n_{1}n_{2} \\
-\lambda_{P}n_{1}n_{2} & 2\lambda_{s}n_{2}^{2}
\end{bmatrix}
\end{equation}
after using the relations in \eqref{eq:GWConditions}. By standard results of linear algebra, this matrix can be diagonalised by a rotation matrix of the form:
\begin{equation}
O=\begin{bmatrix}
\cos\theta &-\sin\theta \\
\sin\theta & \cos\theta
\end{bmatrix}
\end{equation}
where
\begin{equation}
\tan\left(2\theta\right)=\frac{\lambda_{P}n_{1}n_{2}}{\lambda_{s}n_{2}^{2}-\lambda_{H}n_{1}^{2}}.
\end{equation}
We can now write the mass eigenstates:
\begin{equation}
\begin{bmatrix}
           h_{1}\\
           h_{2}
\end{bmatrix}
=O
\begin{bmatrix}
           h\\
           s
\end{bmatrix}
\end{equation}
where we identify $h_{1}$ with the SM higgs. The mass eigenvalues are given by
\begin{equation}
M_{h_{1},h_{2}}^{2}=w^{2}\left(\lambda_{H}n_{1}^{2}+\lambda_{s}n_{2}^{2}\pm\sqrt{\left(\lambda_{H}n_{1}^{2}-\lambda_{s}n_{2}^{2}\right)^{2}+\lambda_{P}^{2}n_{1}^{2}n_{2}^{2}}\right).
\end{equation}
After using the relations in \eqref{eq:flatDirection} to simplify this we obtain
\begin{gather}
\label{eq:h1Mass}
M_{h_{1}}^{2}=\lambda_{P}w^{2}\\
M_{h_{2}}^{2}=0.
\label{eq:massless2}
\end{gather}
We recall that $w$ is a classically flat direction that will be stabilised in \eqref{eq:rewrittenPotential}
and also note that \eqref{eq:massless2} is true only at tree level ({\it cf.} \eqref{eq:h2Mass} below).
We shall take the tree level mass for $h_{1}$ but although $h_{2}$ is massless at tree level it receives sizeble corrections at the one-loop level which we shall calculate at the end of this section.

\medskip

To find the minimum in the quantum theory and lift the classically flat direction $w$, we should now calculate the one loop effective potential for the scalar fields \eqref{eq:shphidef}. A standard calculation in the $\overline{MS}$ scheme leads to the result (see e.g. \cite{Quiros:1999jp} for  a review),
\begin{gather} \nonumber
V_1\left(\phi\right)=\frac{1}{64\pi^{2}}\left(\, \sum_{\rm bosons}n_{i} M_{i}^{4}\left(\phi\right)\left(\log\left(\frac{M_{i}^{2}\left(\phi\right)}{\Lambda_{GW}^{2}}\right)-\frac{3}{2}\right)\right.\\
 -\,\left.\sum_{\rm fermions}n_{i} M_{i}^{4}\left(\phi\right)\left(\log\left(\frac{M_{i}^{2}\left(\phi\right)}{\Lambda_{GW}^{2}}\right)-\frac{3}{2}\right)\right),
\end{gather}
where we have set the RG scale $\mu$ at which the effective potential is computed to be equal to $\Lambda_{GW}$. This allowed us to drop the tree-level $\phi^4$ contribution~\eqref{eq:phi4GW} to the expression for $V_1\left(\phi\right)$.

We also note that since $h_{2}$ is massless at tree level it does not contribute to the effective potential at one loop so the summation runs over $h_{1},\,W,\,Z,\,Z',\,t,\chi_{1},\,\chi_{2}$ with degrees of freedom $n_{i}=1,\,6,\,3,\,3,\,12,\,4,\,4$ respectively. Furthermore, while in principle all SM fermions contribute to the potential, we, as is standard in the literature, account only for the contribution of the top quark (with a factor of three due to colour) as this is the most significant. Since we are working in a theory with no intrinsic masses we can write for all particles: $M^{2}\left(\phi\right)=\frac{M^{2}\phi^{2}}{w^{2}}$ where $M^{2}$ is the observed mass matrix evaluated at $\phi=w$, so we may rewrite the above equation as
\begin{gather}
V_{1}\left(\phi\right)=A\phi^{4}+B\phi^{4}\log\left(\frac{\phi^{2}}{\Lambda_{GW}^{2}}\right)
\label{eq:Veffphi}
\end{gather}
where
\begin{gather}
A=\frac{1}{64\pi^{2}w^{4}}\left(\,\sum_{\rm bosons}n_{i} M_{i}^{4}\left(\log\left(\frac{M_{i}^{2}}{w^{2}}\right)-\frac{3}{2}\right)-\sum_{\rm fermions}n_{i} M_{i}^{4}\left(\log\left(\frac{M_{i}^{2}}{w^{2}}\right)-\frac{3}{2}\right)\right)\\
B=\frac{1}{64\pi^{2}w^{4}}\left(\,\sum_{\rm bosons}n_{i} M_{i}^{4}-\sum_{\rm fermions}n_{i} M_{i}^{4}\right).
\label{eq:Bdef}
\end{gather}

\medskip

Using the expression for the effective potential \eqref{eq:Veffphi} we can now determine the value of the $\phi$ vev $w$,
by solving $(dV_1/d \phi)|_{\phi=w} =0$. 
Doing this we obtain the relationship:
\begin{equation}
\label{eq:GWScaleDef}
\log\left(\frac{w}{\Lambda_{GW}}\right)=-\frac{1}{4}-\frac{A}{2B},
\end{equation}
which determines $w$ in terms of the dimensional transmutaion scale $\Lambda_{GW}$ (which we recall is the scale where the Gildener-Weinberg relations between the scalar couplings \eqref{eq:GWConditions}-\eqref{eq:GWConditions3}
were satisfied).

Using the relation~\eqref{eq:GWScaleDef} allows us to simplify the formula for our potential by removing the dependence on $\Lambda_{GW}$ in favour of the now determined vev $w$, 
\begin{equation}
\label{eq:rewrittenPotential}
V_{1}\left(\phi\right)=B\phi^{4}\left(\log\left(\frac{\phi^{2}}{w^{2}}\right)-\frac{1}{2}\right).
\end{equation}

At the one loop level the mass of $h_{2}$ is given by\footnote{By examining previous relations one can show that $h_{2}$ and $\phi$ turn out to be the same field}
\begin{equation}
\label{eq:h2Mass}
M_{h_{2}}^{2}=\frac{\partial^{2}V}{\partial\phi^{2}}\bigg\rvert_{\phi=w}=8Bw^{2}.
\end{equation}
The value and the sign of constant $B$ is determined by the matter content of the theory (largely by interactions of $\phi$ with the Standard Model fields) via \eqref{eq:Bdef}. In what follows, we will only consider the models for which $B \ge 0$ so that  $M_{h_{2}}^{2}\ge 0$.

\medskip

Finally we end this section with a summary of which parameters are free and which others are determined by the constraints previously listed. Firstly $v_{h}=246$ GeV and $M_{h_{1}}=125$ GeV are known from experiment. $\lambda_{h}$ has a certain value within the Standard Model but it has not been experimentally measured so we shall regard this as undetermined.  We have only one remaining degree of freedom in the scalar sector, once we have picked a value of e.g. $w$ then $\lambda_{P}$ is determined by \eqref{eq:h1Mass} and once $\lambda_{P}$ is determined then the remaining scalar couplings must take their values to satisfy \eqref{eq:GWConditions} (with $n_{1}$ and $n_{2}$ already being determined by the vevs). For our purposes it shall be more convenient to take $\sin\theta$, the mixing angle as our free parameter and determine the scalar couplings and vevs from here.

We shall also take $M_{h_{2}}$ as a free parameter and then $M_{Z'}$ is determined by \eqref{eq:h2Mass}, which in turn determines $g_{D}$ as $M_{Z'}=g_{D}v_{S}$. Finally we have complete freedom in choosing the mass of our fermions, $M_{\chi_{1}},\, M_{\chi_{2}}$ and these in turn shall determine the yukawa couplings $y_{D, 1/2}$. For later convenience we also define $\Delta M_{\chi}$ as the mass splitting between the two fermions and without loss of generality we shall always take $\chi_{1}$ to be the lighter of the two.
%=\sqrt{2}M_{\chi_{1/2}}/v_{S}$.  

\medskip

In summary, the free parameters of our model are $\sin\theta$, $M_{h_{2}}$, $M_{\chi_{1}}$ and $M_{\chi_{2}}$.

\section{Relic density and Experimental Constraints \label{sec:expCon}}

We can consider our model as a model of dark matter, with $\chi_{1}$ and $\chi_{2}$ serving as the dark matter candidates. To calculate the relic density provided by our mode we use MicrOMEGAs \cite{Belanger:2018ccd} with FeynRules \cite{Alloul:2013bka} being used to generate the model file. At the same time we also use MicrOMEGAs to implement several experimental constraints on our model. 

One of the primary constraints on dark sector models comes from direct detection experiments where dark sector particles can scatter off standard model nuclei. This happens in our model due to the mixing between the two scalars.
% and would occur as shown in Fig. \ref{fig:DDExp}. 
This constraint can be done within MicrOMEGAs. 

We also have constraints on the scalar sector of our model. There have been many searches at the LHC for additional light and heavy scalars. So far all such searches have produced null results and so these analyses constrain the valid parameter space of our model. We implement these constraints using the HiggsBounds and HiggsSignals codes \cite{Bechtle:2020pkv,Bechtle:2020uwn}.

\begin{figure}
\begin{subfigure}{.5\textwidth}
\centering
\includegraphics[width=0.95\linewidth,keepaspectratio]{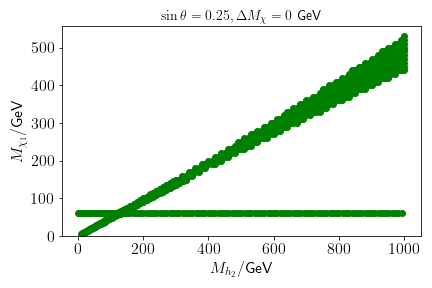}
\caption{The points which correctly produce an acceptable relic abundance for $\sin\theta=0.25,\,\Delta M_{\chi}=0$ GeV.}
\label{fig:DMresonance}
\end{subfigure}%
\hskip 0.3truecm
\begin{subfigure}{.5\textwidth}
\centering
\includegraphics[width=0.95\linewidth,keepaspectratio]{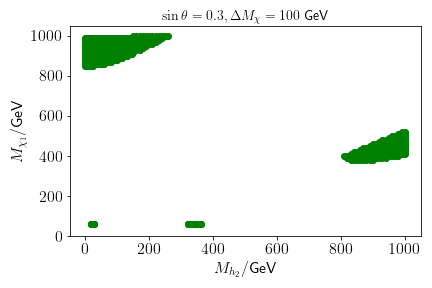}
\caption{The points which produce an acceptable relic abundance for $\sin\theta=0.30,\,\Delta M_{\chi}=100$ GeV.}
\label{fig:DMheavy}
\end{subfigure}
\caption{Areas of our parameter space which do not produce an over-abundance of dark matter.}
\end{figure}

Below we plot the relic density as a function of some of our free parameters and also show some of the constraints coming from direct detection and collider searches. The relic density of the universe has been measured as $\Omega_{DM}h^{2}=0.1200\pm0.0020$ \cite{Workman:2022ynf}. As can be seen in Fig.~\ref{fig:DMresonance}, in order to obtain the correct relic density, we need the mass of our dark fermions to lie in the region around a resonance i.e. $M_{{\chi_{1,2}}}\approx M_{h_{1,2}}/2$, although it should be noted that  the allowed region is not particularly narrow. This near-resonance regime is necessary in order for the dark matter to annihilate sufficiently quickly to not produce an overabundance. An alternative is to have the dark matter sufficiently heavy that the annihilation rate is  enhanced by the phase space, as shown in Fig.~\ref{fig:DMheavy}. Note that the smaller the value of $\sin\theta$ the narrower the resonance, or the larger the mass of the dark fermions should be as the annihilation rate is additionally suppressed.
Here $\Delta M_{\chi}=|M_{\chi_1}-M_{\chi_2}|$ and we always choose $\chi_{1}$ to be the lighter fermion.

\begin{figure}
\begin{subfigure}{.5\textwidth}
\includegraphics[width=0.99\linewidth,keepaspectratio]{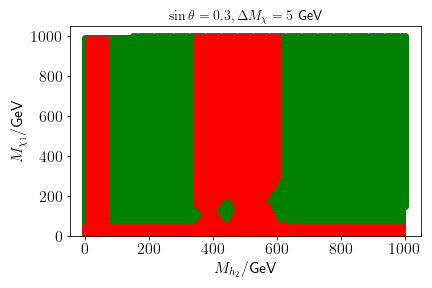}
\caption{Scatter plot of $M_{\chi_{1}}$ against  $M_{h_{2}}$ for $\sin\theta=0.30, \Delta M_{\chi}=5$ GeV with points allowed by constraints from the scalar sector in green and forbidden points in red.}
\label{fig:scalarCon}
\end{subfigure}
\hskip 0.3truecm
\begin{subfigure}{.5\textwidth}
\includegraphics[width=0.99\linewidth,keepaspectratio]{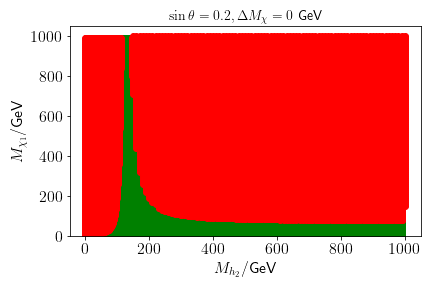}
\caption{Scatter plot of $M_{h_{2}}$ against $M_{\chi_{1}}$ for $\sin\theta=0.20, \Delta M_{\chi}=0$ GeV with points allowed by constraints from the direct detection experiments in green and forbidden points in red.}
\label{fig:DDCon}
\end{subfigure}
\caption{Constraints from collider and direct detection experiments.}
\end{figure}

Hence non-observation of dark matter at the LHC corresponds to an upper bound on the value of $\sin\theta$. Experimental evidence coming from the observed Higgs signal rates requires $\sin\theta<0.44$ independent of the mass of $h_{2}$. There is also a mass-dependent constraint, which requires $\sin\theta\lesssim0.32$ for $M_{h_{2}}\gtrsim200$ GeV and $\sin\theta\lesssim0.2$ for $M_{h_{2}}\gtrsim400$ GeV, mostly coming from restrictions on the NLO corrections to the mass of the W boson (obviously other constraints exist but none as severe as those coming from the W boson mass in our considered parameter range) \cite{Robens:2015gla}. We also in general require $M_{h_{2}}>M_{h_{1}}/2$ to respect bounds coming from the decays of the SM Higgs to invisibles. We show an example plot of the allowed region of parameter space in Fig.~\ref{fig:scalarCon}. The constraints are largely independent of the fermion mass splitting (although there is some effect). 

There are also constraints on the masses of our dark fermions coming from direct detection experiments. Although the fermions do not interact directly with SM quarks/hadrons, they can still interact through the exchange of a mixed scalar, although such diagrams are suppressed by a factor of $\sin\theta$. Such interactions are proportional to $\frac{1}{M_{h_{1}}^{2}}-\frac{1}{M_{h_{2}}^{2}}$ \cite{Bell:2016ekl} and so we require $M_{h_{1}}\approx M_{h_{2}}$ to avoid direct detection constraints. Alternatively we can suppress these diagrams by taking the yukawa coupling of the dark fermions to the scalars to be small i.e. our dark fermions will be light. As one would expect, these constraints become more relaxed for smaller values of $\sin\theta$. These constraints are shown in Fig.~\ref{fig:DDCon} and as for the scalar sector, the constraints are mostly independent of $\Delta M_{\chi}$.  

\section{Theoretical Constraints \label{sec:theCon}}

We shall now examine constraints on the coupling constants coming from vacuum stability, perturbativity and unitarity. From \eqref{eq:rewrittenPotential}, we see that the potential is bounded from below and hence the vacuum is stable if and only if $B\geq0$. 
As pointed out at the end of Section 2, we only consider the models which satisfy this condition and as a result 
we have a positive value of $M_{h_{2}}^2$ which we treat as one of our free parameters. We also require that the vacuum be stable (bounded from below) at tree level which implies
\begin{equation}
\lambda_{P}^{2}<4\lambda_{S}\lambda_{H}\,, \quad \lambda_{H}>0.
\end{equation}
 The requirement of perturbativity simply imples that we have $\lvert g_{i}\rvert<\textrm{const.}$ for all couplings $g_{i}$, i.e. $g_{i}=\lambda_{P},\,\lambda_{H},\,y_{D}\ldots$ \cite{Robens:2015gla}. We choose a constant of $2\pi$ in agreement with \cite{Karam:2016bhq} for the gauge and Yukawa couplings. Due to the difference in loop corrections for scalar couplings we impose the bound  $\lvert \lambda_{i}\rvert<4\pi^{2}$ for the three scalar couplings. We derive and numerically solve the RG equations using SARAH \cite{Staub:2013tta}, and list them in Appendix \ref{appendix:RG}.

Checking the resulting constraints  involves evolving the various coupling constants up to high scales using numerical solutions to the RG equations. Rather than doing it for the entire parameter space we will check these constraints for a selection of bench mark points which we define in the next section. Also it is not necessary for these conditions to hold at arbitrarily high scales (perturbativity and vacuum stability are not absolute requirements in any case) as there may be new physics which arises at some higher scale which then contributes in such a way to e.g. stabilise the vacuum. Hence when we numerically check these constraints we only require them to hold up to $\Lambda_{GW}$ (defined by Eq.~\ref{eq:GWScaleDef}) and then give the higher scale at which they are violated, see Table~\ref{Table:benchPoints}.

We now consider constraints from perturbative unitarity for our theory. A partial wave expansion for a scattering amplitude gives
\begin{equation} 
\mathcal{M}\left(s,\theta\right)=16\pi\sum_{J=0}^{\infty}\left(2J+1\right)A_{J}\left(s\right)P_{K}\left(\cos\theta\right)
\end{equation}
where $P_{J}$ are the Legendre polynomials and $A_{J}$ are the partial wave amplitudes. Unitarity then imposes the bound that $\lvert Re A_{0}\rvert<\frac{1}{2}$. We consider the tree level amplitudes for the processes: $Z'_{L}Z'_{L}\rightarrow Z'_{L}Z'_{L}, h_{1}h_{1}\rightarrow h_{1}h_{1}, h_{2}h_{2}\rightarrow h_{2}h_{2}$. We use FeynArts \cite{Hahn:2000kx} and FeynCalc \cite{Shtabovenko:2020gxv} to aid in the computation of the amplitudes. For the scalar process, the only relevant diagram at high energy is the four-point interaction (all others are suppressed by $\sim\frac{1}{s}$ due to internal  propagators) and so the demand for perturbative unitarity simply imposes
\begin{gather}
\label{eq:sc_un_1}
\frac{6}{16\pi}\left(\lambda_{H}\cos^{4}\theta-\lambda_{P}\sin^{2}\theta\cos^{2}\theta+\lambda_{S}\sin^{4}\theta\right)<\frac{1}{2} \\
\label{eq:sc_un_2}
\frac{6}{16\pi}\left(\lambda_{H}\sin^{4}\theta-\lambda_{P}\sin^{2}\theta\cos^{2}\theta+\lambda_{S}\cos^{4}\theta\right)<\frac{1}{2}.
\end{gather}
For the vector boson scattering we obtain
\begin{gather}
\mathcal{M}=-4\frac{g_{D}^{2}v_{s}^{2}\sin^{2}\theta}{M_{Z'}^{4}}\left(\frac{\left(s-2M_{Z'}^{2}\right)^{2}}{s-M_{h_{1}}^{2}}+\frac{\left(t-2M_{Z'}^{2}\right)^{2}}{t-M_{h_{1}}^{2}}+\frac{\left(u-2M_{Z'}^{2}\right)^{2}}{u-M_{h_{1}}^{2}}\right)-\\
4\frac{g_{D}^{2}v_{s}^{2}}{M_{Z'}^{4}}\cos^{2}\theta\left(\frac{\left(s-2M_{Z'}^{2}\right)^{2}}{s-M_{h_{2}}^{2}}+\frac{\left(t-2M_{Z'}^{2}\right)^{2}}{t-M_{h_{2}}^{2}}+\frac{\left(u-2M_{Z'}^{2}\right)^{2}}{u-M_{h_{2}}^{2}}\right)\\
\approx -4\frac{\sin^{2}\theta}{M_{Z'}^{2}}\left(s+t+u+3M_{h_{1}}^{2}\right)-4\frac{\cos^{2}\theta}{M_{Z'}^{2}}\left(s+t+u+3M_{h_{2}}^{2}\right)\\
\approx -4\sin^{2}\theta\left(4+3\frac{M_{h_{1}}^{2}}{M_{Z'}^{2}}\right)-4\cos^{2}\theta\left(4+3\frac{M_{h_{2}}^{2}}{M_{Z'}^{2}}\right).
\end{gather}
Hence we require
\begin{equation}
\label{eq:vec_un}
4\sin^{2}\theta\left(4+3\frac{M_{h_{1}}^{2}}{M_{Z'}^{2}}\right)+4\cos^{2}\theta\left(4+3\frac{M_{h_{2}}^{2}}{M_{Z'}^{2}}\right)<8\pi.
\end{equation}
Equations \eqref{eq:sc_un_1}, \eqref{eq:sc_un_2}, \eqref{eq:vec_un}
summarise the unitarity constraints that we require to hold for our model.

\medskip

\section{Phase transition and Gravitational Wave signal\label{sec:gravWaves}}

To discuss the phase transition and gravitational waves we must first compute the one-loop effective potential at finite temperature. It is known that at one loop level, the potential factorises into the zero temperature potential (which we have already calculated) plus thermal corrections. The thermal corrections are given by\footnote{This can be calculated using exactly the same diagrams as the zero-temperature potential but now using the Feynman rules for a theory at finite temperature, see \cite{Quiros:1999jp} for a review}
\begin{gather}
V_{T}=\frac{T^{4}}{2\pi^{2}}\left(\sum_{bosons} n_{i}J_{B}\left(\frac{M_{i}^{2}\left(\phi\right)}{T^{2}}\right)-\sum_{fermions} n_{i}J_{F}\left(\frac{M_{i}^{2}\left(\phi\right)}{T^{2}}\right)\right)
\end{gather}
where the functions $J_{B/F}$ are defined by 
\begin{equation}
J_{B/F}\left(x^{2}\right)=\int_{0}^{\infty}\,dy\,y^{2}\log\left(1\mp e^{-\sqrt{x^{2}+y^{2}}}\right).
\end{equation}

\begin{figure}
\centerline{\includegraphics[width=0.45\textwidth,keepaspectratio]{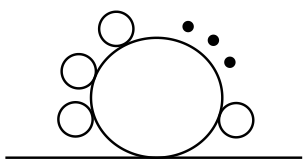}}
\caption{An example of a daisy diagram \cite{Quiros:1999jp}  with a scalar field appearing in the outside bubbles. This is then resummed to all orders (the outside series of bubbles) to obtain the thermal mass correction.}
\label{fig:daisyDiagram}
\end{figure}
Finally to go beyond the simple one-loop expressions for the effective potential we were using until now, we now add the resummed contributions of the so-called `daisy diagrams' (shown in Fig.~\ref{fig:daisyDiagram}) to improve the validity of perturbation theory.
This was first done for the Standard Model in \cite{Carrington:1991hz,Arnold:1992rz}, and for our model we have
\begin{gather}
\label{eq:daisy}
V_{daisy}=\frac{T}{12\pi}\sum_{bosons} n_{i} \left(M_{i}^{3}\left(\phi\right)-\left(M_{i}^{2}+\Pi_{i}\left(\phi,T\right)\right)^{\frac{3}{2}}\right)
\end{gather}
where $\Pi_{i}\left(\phi,T\right)$ is the thermal mass correction. Note that, to leading order, fermions do not receive a thermal mass and so do not contribute to the daisy potential \eqref{eq:daisy} and also that it is only the longitudinal mode of the gauge bosons which receives a thermal mass, so the relevant degrees of freedom should be divided by three. The thermal masses are given by:
\begin{gather}
\label{eq:scalarThermalMass}
\Pi_{h/s}=\begin{bmatrix}
T^{2}\left(\frac{\lambda_{H}}{4}+\frac{\lambda_{P}}{24}+\frac{g'^{2}}{16}+\frac{3g_{W}^{2}}{16}+\frac{y_{t}^{2}}{4}\right) & 0 \\
0 & T^{2}\left(\frac{\lambda_{S}}{4}+\frac{\lambda_{P}}{24}+\frac{g_{D}^{2}}{4}+\frac{y_{\chi_{1}}^{2}}{12}+\frac{y_{\chi_{2}}^{2}}{12}\right) 
\end{bmatrix} \\
\Pi_{Z'}=\frac{g_{D}^{2}T^{2}}{3} \\
\Pi_{W}=\frac{11}{6}g_{W}^{2}T^{2} \\
\Pi_{Z}=\frac{11}{6} g_{W}^{2}T^{2}
\end{gather}
where the results for the $W$ and $Z$ bosons were taken from \cite{Carrington:1991hz}. Hence the full one-loop effective potential is given by 
\begin{equation}
V\left(\phi,T\right)=V_{1}\left(\phi\right)+V_{T}\left(\phi,T\right)+V_{daisy}\left(\phi,T\right)
\end{equation}
Note that to determine the mass of the scalars at finite temperature one must add Eqs.~\ref{eq:massMatrix} and \ref{eq:scalarThermalMass} before finding the eigenvalues. It is known that in order to generate matter-antimatter asymmetry we must have a strongly first order electroweak phase transition. The order parameter for these transitions is given by the ratio ${\phi_{c}}{/T_{c}}$, where the critical temperature, $T_{c}$ , and the critical field strength, $\phi_{c}$ are defined by
\begin{gather}
V\left(\phi_{c},T_{c}\right)=0\\
\partial_\phi V\left(\phi_{c},T_{c}\right)=0,
\end{gather}
i.e. $\phi_{c}$ is a local minimum of the potential that is degenerate with the minimum at the origin at $T_{c}$. To have a strongly first order phase transition we then require ${\phi_{c}}/{T_{c}}\gtrsim1$. The numerical calculation of the order parameters and of various parameters associated with the gravitational wave signal becomes quite slow and so in this section rather than completing a full exploration of the phase space we choose several benchmark points consistent with the constraints from Sections \ref{sec:expCon},\ref{sec:theCon} and compute the order parameters and gravitational wave signal. Our benchmark points are listed in Table~\ref{Table:benchPoints}.

\begin{table}
\centering
\scalebox{0.8}{
\begin{tabular}{|c|c|c|c|c|c|c|c|c|}
\hline
 &  $\sin\theta$                           &$ M_{h_{2}}     $                     & $M_{\chi_{1}} $ & $M_{\chi_{2}}$   & $\Omega h^{2}$           & $\Lambda_{\textrm{unit}}    $        & $ \Lambda_{\textrm{pert.}}$ &$ \Lambda_{\textrm{stab.}}             $       \\ 
\hline
BP1 &              0.30                                &         151\, \textrm{GeV}  &   59.5\, \textrm{GeV}   &   59.5\, \textrm{GeV}     &       0.070         &  $   2.2\times10^{9}\, \textrm{GeV}      $       &     $    >10^{16}\, \textrm{GeV}  $   &  $ 9.3\times10^{4}\, \textrm{GeV}  $ \\
\hline
BP2 &                  0.10                            &       320\, \textrm{GeV}    &      150\, \textrm{GeV}  &       155\, \textrm{GeV}  &     0.078     &    $   3.0\times10^{15}\, \textrm{GeV}    $     &      $       >10^{19}\, \textrm{GeV}   $    &  $ 3.2\times10^{5}\, \textrm{GeV} $ \\
\hline 
BP3 &                     0.40                        &    121\, \textrm{GeV}        &    591\, \textrm{GeV}         &   592\, \textrm{GeV}      &       0.118 &  $  6.3\times10^{8}\, \textrm{GeV}     $           &    $    >10^{10}\, \textrm{GeV}   $    &  $ 8.9\times10^{4}\, \textrm{GeV} $ \\
\hline
BP4 &                       0.20                    &      331\, \textrm{GeV}          &      61\, \textrm{GeV}      &   161\, \textrm{GeV}     &      0.077    &       $   1.8\times10^{6}\, \textrm{GeV}     $   &      $   >10^{12}\, \textrm{GeV}  $   &   $ 2.0\times10^{5}\, \textrm{GeV}  $ \\
\hline
BP5 &                         0.30                  &     120\, \textrm{GeV}   &      901\, \textrm{GeV}      &     1001\, \textrm{GeV}        &       0.118    &    $   5.8\times10^{6}\, \textrm{GeV}    $     &        $   >10^{9}\, \textrm{GeV}      $     &  $ 1.2\times10^{5}\, \textrm{GeV}  $ \\
\hline
\end{tabular}
}
\caption{Table showing our selection of benchmark points. The $\Lambda$ show the scale at which we violate perturbativity, perturbative unitarity and vacuum stability respectively. All chosen points also obey the experimental constraints coming from collider searches and direct detection experiments. Note that due to numerical issues in the software we were unable to determine the exact scale at which perturbativity is violated for most of our benchmark points and so we indicate the maximum scale we were able to check. }
\label{Table:benchPoints}
\end{table}

It is well known that a strongly first order phase transition will produce a gravitational wave signal. Here we calculate this signal and examine the possibility of detection at both present detectors (LIGO, VIRGO etc.) and future detectors (LISA, DECIGO etc.).

A first order phase transition occurs when there is a potential barrier between a false minimum (usually at $\phi=0$) and a true minimum. When this occurs the transition happens as bubbles of true vacuum nucleate in the false vacuum. A gravitational wave signal is produced by three different mechanisms, as reviewed in \cite{Caprini:2015zlo}): collisions between bubbles, sound waves in the plasma, and magnetohydrodynamic turbulence. 

Before going on to calculate the signal we briefly outline some bubble nucleation theory necessary for our calculation. The vacuum at $\phi=0$ only becomes metastable at temperatures $T<T_{c}$, however if the barrier is sufficiently high then the tunnelling rate may remain very small even for temperatures much below the critical temperature. Hence it is conventional to also define the \textit{nucleation temperature} at which the probability of one bubble nucleating in one horizon volume is approximately one. The theory of such transitions and bubble nucleation was first addressed in \cite{Coleman:1977py,Callan:1977pt} where it was shown that the decay rate was given by
\begin{equation}
\frac{\Gamma}{V}=Ae^{-S_{4}}
\end{equation}
where the left-hand side is the decay rate per unit volume and on the right-hand side we have $A$ which is a ratio of determinants of quadratic fluctuation operators around the bubble solution. $S_{4}$ is the action computed on the field profile, $\phi$, satisfying the differential equation
\begin{equation}
\frac{d^{2}\phi}{d\rho^{2}}+\frac{3}{\rho}\frac{d\phi}{d\rho}=V'\left(\phi\right)
\end{equation}
which is the Euler-Lagrange equation for a field in four dimensions with an $O\left(4\right)$ symmetry, $\phi({\bf x}, t)=\phi(\rho)$,
and the boundary conditions,
\begin{equation}
\label{eq:bcb}
\lim_{\rho\to\infty} \phi(\rho) = 0\,, \quad \partial_\rho \phi(0) =0.
\end{equation}
The solution to this classical problem corresponds to the four-dimensional bubble or bounce configuration.

It was shown in \cite{Linde:1981zj} that when working in a theory at finite temperature this four-dimensional approach should be modified to the effectively three-dimensional set-up,
\begin{equation}
\frac{\Gamma}{V}=Ae^{-S_{3}/T}
\end{equation}
and the field profile $\phi$ should satisfy
\begin{equation}
\frac{d^{2}\phi}{d\rho^{2}}+\frac{2}{\rho}\frac{d\phi}{d\rho}=V'\left(\phi\right)
\end{equation}
with the same boundary conditions \eqref{eq:bcb}. 
At finite temperature, due to the periodicity of the imaginary time-dimension $0\le \tau \le 1/T$ in the Matsubara formalism, we essentially work in a three-dimensional theory\footnote{The D-dimensional action is given by $S_{D}=\int d\rho d\Omega_{D} \, \rho^{D-1}\left[\left(\frac{d\phi}{d\rho}\right)^{2}+V\left(\phi\right)\right]$ where $\Omega_{D}$ is an integral over the surface of a D-dimensional sphere.}. Note that it is not possible, in general, to analytically calculate the value of the prefactor $A$; instead it is common practice in the literature to take it as $\mathcal{O}\left(T^{4}\right)$ on dimensional grounds.

It was shown in \cite{Apreda:2001us} that the nucleation temperature, $T_{N}$, is given by solving the equation $S_{3}\left(T_{N}\right)/T_{N}\approx140$. 
To describe  the gravitational wave spectrum resulting from the first-order phase transition detailed above, 
it is conventional to define two more parameters $\alpha$ and $\beta$, in addition to the nucleation temperature $T_N$,
 that characterise the phase transition:
\begin{gather}
\alpha=\frac{1}{\rho_{rad}(T_{N})}\left(\Delta V(T_{N})-T_{N}\frac{d \Delta V}{dT}\bigg\rvert_{T=T_{N}}\right) \\
\frac{\beta}{H_{*}}=T_{N}\frac{d(S_{3}/T)}{dT}\bigg\rvert_{T=T_{N}}
\end{gather}
where $H_{*}$ is the Hubble constant at the time of nucleation, $\rho_{rad}$ is the radiation energy density\footnote{$\rho_{rad}(T_{N})=g_{*}\pi^{2}T_{N}^{4}/30$ where $g_{*}$(=117.75 for our model) is the number of relativistic degrees in the plasma at $T_{N}$.} and $\Delta V(T)=V(0,T)-V(v(T),T)$ where $v(T)$ is the global minimum of the potential at temperature $T$. The gravitational wave energy density, $\Omega_{GW}$, as a function of frequency, $f$, is then given by the sum of the three production modes \cite{Caprini:2015zlo}
\begin{gather}
\Omega_{Coll}\,h^{2}=1.67\times10^{-5}\left(\frac{H_{*}}{\beta}\right)^{2}\left(\frac{\kappa\alpha}{1+\alpha}\right)^{2}\left(\frac{100}{g_{*}}\right)^{\frac{1}{3}}\left(\frac{0.11v_{w}^{3}}{0.42+v_{w}^{2}}\right)S_{Coll}\left(f\right)\\
\Omega_{SW}\,h^{2}=2.65\times10^{-6}\left(\frac{H_{*}}{\beta}\right)\left(\frac{\kappa_{v}\alpha}{1+\alpha}\right)^{2}\left(\frac{100}{g_{*}}\right)^{\frac{1}{3}}v_{w}\,S_{SW}\left(f\right)\\
\Omega_{MHD}\,h^{2}=3.35\times10^{-4}\left(\frac{H_{*}}{\beta}\right)\left(\frac{\kappa_{MHD}\alpha}{1+\alpha}\right)^{\frac{3}{2}}\left(\frac{100}{g_{*}}\right)^{\frac{1}{3}}v_{w}\,S_{MHD}\left(f\right)
\end{gather}
where  $S\left(f\right)$ are the known functions parametrising the dependence on frequency (i.e. determining the shape of the curve as a function of frequency), $v_{w}$ is the velocity of the bubble walls and the $\kappa$-parameters denote the fraction of latent heat that is transformed into sources relevant to each production mode. The precise contribution of the different sources of gravitational waves and formulae for $\kappa$ depend on the dynamics of the bubble walls, see \cite{Caprini:2015zlo} for more details. To determine which regime we lie in we must determine whether the bubble walls are relativistic and whether they `runaway'($\gamma\rightarrow\infty$). 

We do not expect runaway walls as our $Z'$ bosons become massive during the transition and it is known that one should not expect runaway bubbles for a transition where gauge bosons gain a mass \cite{Bodeker:2017cim,Croon:2018erz}. A strongly first order phase transition is expected to give highly relativistic bubble walls and so we take $v_{w}=1$. The exact formulae for the $S(f)$ and $\kappa$ (for our regime) are given in Appendix~\ref{appendix:GW}. In this regime the contribution from collision of bubble walls is negligible so we do not include this in our calculations.

It should be noted that these formulae are intended to give only a first approximation to the gravitational wave signal. For more precise calculations and recent work on computing gravitational wave spectra we refer the reader to e.g.\ \cite{Ellis:2020awk,Lewicki:2020azd,Caldwell:2022qsj,Brdar:2018num}.

\begin{figure}
\centerline{\includegraphics[width=0.9\textwidth,keepaspectratio]{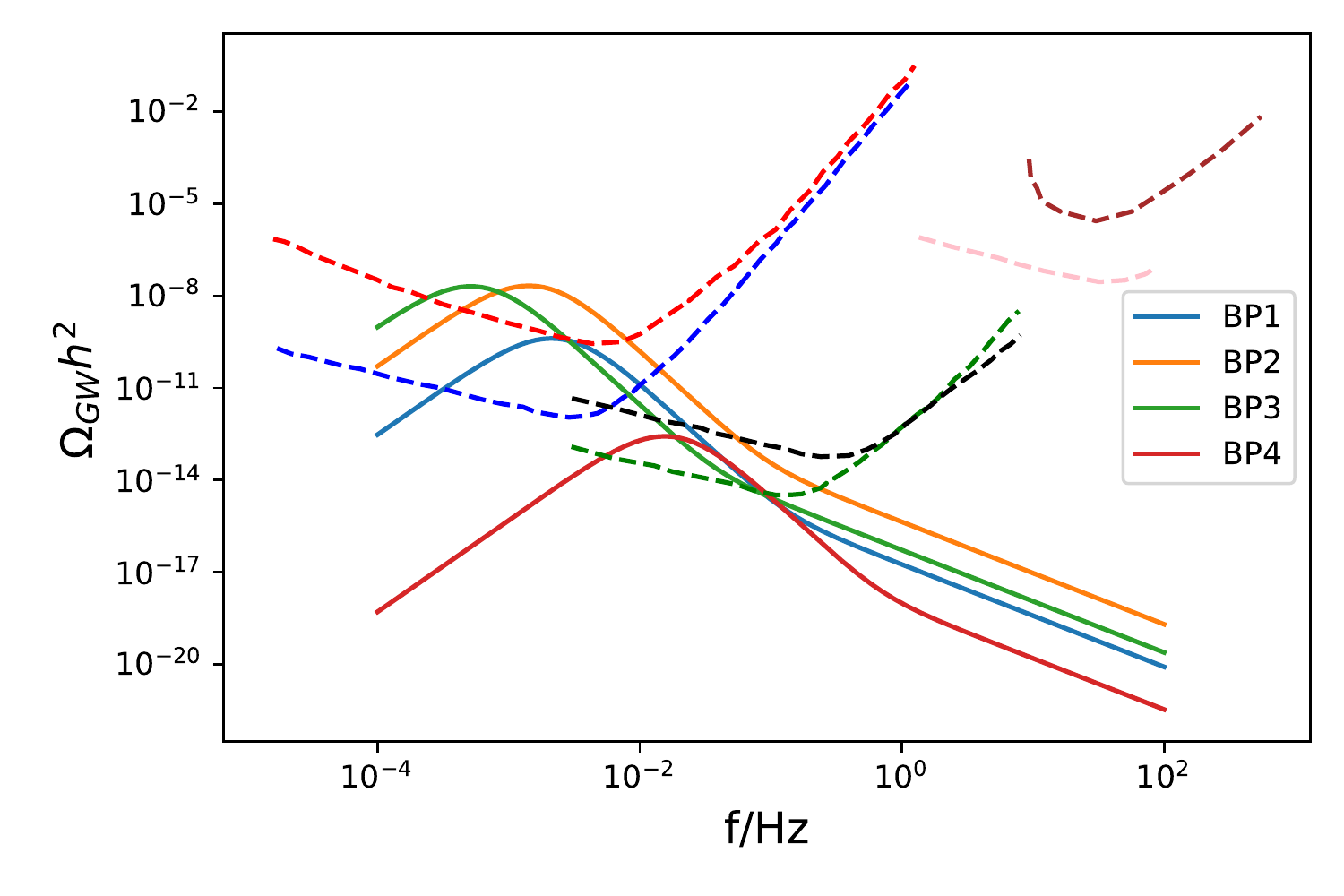}}
\caption{A plot showing the energy density of gravitational waves for the first four benchmark points. The dashed lines represent the sensitivities of current and future gravitational wave detectors: LISA (blue), eLISA (red), BBO (green), DECIGO (black), Einstein Telescope (pink) and aLIGO (brown).}
\label{fig:gravWaves}
\end{figure}

We calculate the bubble profile and the action on the profile using BubbleProfiler \cite{Athron:2019nbd}. The nucleation temperatures and parameters $\alpha,\beta$ are shown in Table~\ref{Table:gravPars} for the benchmark points. The gravitational wave profiles are then plotted in Fig.~\ref{fig:gravWaves} along with the sensitivities of current and planned gravitational wave detectors.

As can be seen from the figure, the gravitational waves produced by our model have too low a frequency to probed by aLIGO but would be probed by the next generation of space-based gravitational wave detectors such as LISA, DECIGO and BBO.

\begin{table}
\centering
\scalebox{0.9}{
\begin{tabular}{|c|c|c|c|c|c|}
\hline
        &$ T_{\textrm{c}} $                & $\phi_{\textrm{c}} $      & $T_{\textrm{N}}$   & $\alpha$      & $\frac{\beta}{H_{*}}        $          \\ 
\hline
BP1 &    221\, \textrm{GeV}             &  750\, \textrm{GeV}         & 84.2\,\textrm{GeV} &  0.547              &    129                                     \\
\hline
BP2 &    622\, \textrm{GeV}             &  2313\, \textrm{GeV}      & 115.2\,\textrm{GeV}   &    5.70           &     63.5                                     \\
\hline 
BP3 &    129\, \textrm{GeV}             &       586\, \textrm{GeV}      &  30.8\,\textrm{GeV}   &   10.1        &      85.4                            \\
\hline  
BP4 &    449\, \textrm{GeV}           &     1150\, \textrm{GeV}         &    273.9\,\textrm{GeV}      &     0.0698      &     290                         \\
\hline
BP5 &   152\, \textrm{GeV}         &         802\, \textrm{GeV}     &             <10 \textrm{GeV}                &                -                  &                        -                     \\
\hline
\end{tabular}
}
\caption{Table showing the value of various parameters which are relevant to gravitational waves for our benchmark points. as can be seen from our values of $\phi_{c}, T_{c}$, all of our benchmark points leads to a strongly first order phase transition. Note that for the fifth benchmark point the nucleation temperature is very low and our software encounters problems in this area. Hence we were unable to determine the exact nucleation temperature (it may be that the model does not nucleate in this region of parameter space) and so we do not determine the gravitational wave spectrum for this point.}
\label{Table:gravPars}
\end{table}

\section{Conclusion \label{sec:conc}}

We have shown in this paper that a classically scale invariant model can evade all current theoretical and experimental constraints and still account for some or all of the observed dark matter abundance of the universe. Such a model is quite an attractive prospect as it is a relatively minimal model which can solve several problems of the Standard model, primarily dark matter and the hierarchy problem (see footnote \ref{footnote:HPDebate}), while also producing a gravitational wave signal which would be observable at the next generation of detectors.

In the context of a minimal model presented here, we have not addressed the question of matter-anti-matter asymmetry. One scenario considered in the literature in classically scale-invariant settings~\cite{Khoze:2013oga,Khoze:2016zfi} is to use a version of leptogenesis via sterile neutrino oscillations~\cite{Akhmedov:1998qx,Drewes:2012ma}, though this would require an extension of our minimal model.

\section*{Acknowledgements}

Research of VVK is supported by the UK Science and Technology Facilities Council (STFC) under grant ST/P001246/1 and DLM acknowledges an STFC studentship.

\newpage

\startappendix
\Appendix{A. RG Equations \label{appendix:RG}}

{\allowdisplaybreaks  \begin{align} 
\beta_{g_{D}}^{(1)} & =  
g_{D}^{3}\\ 
\end{align}
{\allowdisplaybreaks  \begin{align} 
\beta_{\lambda_H}^{(1)} & =  
+\frac{27}{200} g'^{4} +\frac{9}{20} g'^{2} g_{W}^{2} +\frac{9}{8} g_{W}^{4} -\frac{9}{5} g'^{2} \lambda_{H}-9 g_{W}^{2} \lambda_{H} +24 \lambda_{H}^{2} +\lambda_{P}^{2}+12 \lambda_{H} \mbox{Tr}\Big({Y_d  Y_{d}^{\dagger}}\Big) \nonumber \\ 
 &+4 \lambda_{H} \mbox{Tr}\Big({Y_e  Y_{e}^{\dagger}}\Big) +12 \lambda_H \mbox{Tr}\Big({Y_u  Y_{u}^{\dagger}}\Big) -6 \mbox{Tr}\Big({Y_d  Y_{d}^{\dagger}  Y_d  Y_{d}^{\dagger}}\Big) -2 \mbox{Tr}\Big({Y_e  Y_{e}^{\dagger}  Y_e  Y_{e}^{\dagger}}\Big) -6 \mbox{Tr}\Big({Y_u  Y_{u}^{\dagger}  Y_u  Y_{u}^{\dagger}}\Big) \\ 
\beta_{\lambda_{P}}^{(1)} & =  
\frac{1}{10} \lambda_{P} \Big(-9 g'^{2} -45 g_{W}^{2} -60 g_{D}^{2}+120 \lambda_{H} -40 \lambda_{P} +80 \lambda_{S} +20 |y_{2,D}|^2 +20 |y_{1,D}|^2 +60 \mbox{Tr}\Big({Y_d  Y_{d}^{\dagger}}\Big) \nonumber \\ 
 &+20 \mbox{Tr}\Big({Y_e  Y_{e}^{\dagger}}\Big) +60 \mbox{Tr}\Big({Y_u  Y_{u}^{\dagger}}\Big) \Big)\\ 
\beta_{\lambda_{S}}^{(1)} & =  
2 \Big(10 \lambda_{S}^{2}  + 2 \lambda_{S} |y_{2,D}|^2  + 2 \lambda_{S} |y_{1,D}|^2  + 3 g_{D}^{4}  -6 g_{D}^{2} \lambda_{S}- |y_{2,D}|^4  - |y_{1,D}|^4  + \lambda_{P}^{2}\Big)\\ 
\end{align}} 
{\allowdisplaybreaks  \begin{align} 
\beta_{y_{1,D}}^{(1)} & =  
\frac{1}{2} y_{1,D} \Big(2 |y_{2,D}|^2  -3 g_{D}^{2}+ 4 |y_{1,D}|^2 \Big)\\ 
\beta_{y_{2,D}}^{(1)} & =  
\frac{1}{2} y_{2,D} \Big(2 |y_{1,D}|^2  -3 g_{D}^{2} + 4 |y_{2,D}|^2 \Big)\\ 
\end{align}} 

\Appendix{B. Gravitational wave formulae \label{appendix:GW}}

All formulae here are taken from \cite{Caprini:2015zlo}. The remaining parameters required for calculation of the gravitational wave spectrum are given below. Firstly we begin with the frequency dependence of the sound wave production.
\begin{equation}
S_{SW}(f)=\left(\frac{f}{f_{SW}}\right)^{3}\left(\frac{7}{4+3\left(\frac{f}{f_{SW}}\right)^{2}}\right)^{\frac{7}{2}}
\end{equation}
where
\begin{equation}
f_{SW}=1.9\times10^{-2}\,\textrm{mHz}\,\left(\frac{1}{v_{w}}\right)\left(\frac{\beta}{H_{*}}\right)\left(\frac{T_{*}}{100\,\textrm{GeV}}\right)\left(\frac{g_{*}}{100}\right)^{\frac{1}{6}}
\end{equation}

The frequency dependence of the gravitational wave production by turbulence is given by
\begin{equation}
S_{turb}(f)=\frac{\left(\frac{f}{f_{turb}}\right)^{3}}{\left(1+\frac{f}{f_{turb}}\right)^{\frac{11}{3}}\left(1+\frac{8\pi f}{h_{*}}\right)} 
\end{equation}
where
\begin{gather}
f_{turb}=2.7\times10^{-2}\,\textrm{mHz}\,\left(\frac{1}{v_{w}}\right)\left(\frac{\beta}{H_{*}}\right)\left(\frac{T_{*}}{100\,\textrm{GeV}}\right)\left(\frac{g_{*}}{100}\right)^{\frac{1}{6}} \\
h_{*}=16.5\times10^{-3} \,\textrm{mHz}\, \left(\frac{T_{*}}{100\,\textrm{GeV}}\right)\left(\frac{g_{*}}{100}\right)^{\frac{1}{6}} 
\end{gather}

The efficiencies of the two processes are given by
\begin{gather}
\kappa_{v}=\frac{\alpha}{0.73+0.083\sqrt{\alpha}+\alpha} \\
\kappa_{turb}=0.05\kappa_{v}
\end{gather}

\newpage

\bibliographystyle{inspire}
\bibliography{main}

\clearpage

\include{variables}

\end{document}